\documentstyle[psfig]{elsart}

\begin{document} 
\begin{frontmatter}

\title{On the Spectrum of Ultrahigh Energy Cosmic Rays and the Gamma Ray 
Burst Origin Hypothesis}
\author{S.T. Scully\thanksref{nsf}}
\author{F.W. Stecker}
\address{NASA Goddard Space Flight Center, Greenbelt, MD 20771}
\thanks[nsf]{National Academy of Sciences-National Research Council Postdoctoral Resident Research Associate}
\begin{abstract}               
  
  It has been suggested that cosmological $\gamma$-ray bursts (GRBs) can
  produce the observed flux of cosmic rays at the highest energies.
  However, recent studies of $\gamma$-ray bursts indicate
  that their redshift distribution likely follows the average
  star formation rate of the universe and that GRBs were more numerous at 
  high redshifts. As a consequence, we show that photomeson production
  energy losses suffered by ultrahigh energy cosmic rays coming from GRBs 
  would produce too sharp a spectral energy cutoff to be consistent 
  with the air shower data. Futhermore, we show that cosmolgical GRBs
  fail to supply the energy input required to account for the cosmic ray
  flux above $10^{19}$ eV  by a factor of 100-1000.

\end{abstract}

\end{frontmatter}

\section{Introduction}

Ultra-high energy cosmic rays (UHECRs) are among the most energetic
form of radiation ever detected, with energies apparently reaching
beyond 10$^{20}$eV.  Cosmic rays with ultrahigh energies were
first detected over 35 years ago by the Volcano Ranch experiment
\cite{vol63}.  With typical event rates of one per km$^2$ per century,
subsequent experiments including the Fly's Eye \cite{fly94} and AGASA
\cite{agasa95}\cite{agasa98} experiments have observed only a double
digit number of events.
However, in spite of the paucity of data, a surprising characteristic 
in the UHECR 
spectrum has been observed, {\it viz.}, its apparent smooth continuation 
beyond $10^{20}$ eV.  UHECRs appear to be
distributed isotropically, indicating an extragalactic origin. Protons
traveling from intergalactic distances should experience energy losses
owing to photopion production interactions with the 2.7K cosmic
background radiation.  For a uniform distribution of protons in the
universe, even neglecting redshift effects and cosmological source
evolution, these interactions would be expected to produce a cutoff in
the observed spectrum of UHECRs at $\sim 6 \times 10^{19}$ eV, commonly
referred to the Greisen-Zatsepin-Kuzmin (GZK)
cutoff\cite{gzk1}\cite{gzk2}.  This is because of the strong energy
dependence in the energy loss rate with energy from photomeson
production interactions. As a result, above $\sim 10^{20}$ eV, only
protons from within the relatively nearby distance of $\le 100$ Mpc
will survive at these energies to reach the Earth \cite{stk68}.  The
present data indicate, however, a flattening of the spectrum above
$\sim 10^{19}$ eV\cite{bir93} and no obvious cutoff out to an energy of 
$\sim3\times 10^{20}$ eV.

Two classes of theories have emerged to explain this
phenomenon.  In one class of theories, ``top down'', a small branching ratio
into high energy nucleons would result from the decays of unstable or
meta-stable supermassive particles originating at the grand unified scale
or in topological defects produced at that scale. (Most of the energy from 
these decays goes into the production of pions which, in turn, decay to 
produce much larger numbers of neutrinos, photons and electrons.) 

In the other class of theories, ``bottom up'', protons achieve their high 
energies through conventional electromagnetic acceleration processes. Only a 
few astrophysical sources are presently considered to have the possibility 
to accelerate particles to ultrahigh energies of the order of $10^{20}$ eV.  
Suggested possible acceleration sites include AGN jets and hot spots at the 
ends of radio galaxy jets.  For reviews, see Ref. \cite{yos98}.

Among other origin hypotheses, it has been suggested 
that $\gamma$-ray bursts (GRBs) can produce the observed flux of
UHECRs\cite{wax95a}.  In such scenarios, it has been speculated 
that UHECRs would be emitted by GRBs fireballs with roughly 
the same amount of total emitted energy as the
observed $\gamma$-rays in the keV to MeV range. Based on
this assumption, and also assuming that the GRB rate is 
independent of redshift, it has been argued that the flux and spectrum of 
UHECRs with energies above $10^{19}$ eV (a putative extragalactic component
which could account for the flattening and possible composition change
above $10^{19}$ eV\cite{bir93}) is consistent with their production by GRBs 
\cite{wax95b}.

In recent years, X-ray, optical, and radio afterglows of about a dozen
GRBs have been detected leading to the subsequent identification of the host
galaxies of these objects and consequently, their redshifts. The regions
in the host galaxies where the bursts are located have also been identified
as regions of very active ongoing star formation.
One of us \cite{stk00} has therefore argued that a more appropriate redshift
distribution to take for GRBs is that of the average star
formation rate. One consequence of most GRBs being at moderate to high
redshifts, is that the GZK effect in the UHECR spectrum produced by GRBs 
would materialize at lower energies and be more pronounced, in stark
contradiction to the data.  In this Letter, we present a
detailed calculation of the expected UHECR spectrum which would result
from GRBs with a more realistic redshift distribution, one
following the star formation rate.

\section{Gamma Ray Burst Redshift Evolution}

To date, some 15 GRBs afterglows of long duration bursts have been detected 
with a subsequent identification of their host galaxies.  At least 14 of the 
15 are at moderate to high redshifts with the highest one (GRB000131) lying 
at a redshift of 4.50\cite{and00},\cite{gal97}.\footnote{The origin of 
GRB980425 is somewhat controversial; a possible
X-ray source\cite{pia99} and an unusual nearby Type Ic
supernova\cite{gal97} have both been put forward as candidates.  Taking the
supernova identification gives an energy release which is 
orders of magnitude smaller 
than that for a typical cosmological GRB.}
Redshift data on GRB host galaxies 
are presently available only for long duration bursts; the short
duration bursts have not as yet been identified with
any astronomical sources, however, their source-count distribution is also
consistent with a cosmological origin \cite{pac99}.

The host galaxies of GRBs appear to be sites of active star
formation\cite{rei98}\cite{cl98}.  The colors and morphological types of these
galaxies are also consistent with active star formation\cite{kul98} as
is the detection of Ly$\alpha$ and [OII] in several of the host
galaxies\cite{met97}\cite{blm98}.  Further evidence 
suggests that bursts themselves are directly associated with star
forming regions within their host galaxies; their positions
correspond to regions having significant hydrogen column densities
with evidence of dust extinction\cite{rei98}\cite{kul98}. 
A good argument in favor of strong
redshift evolution for GRBs is made in Ref.\cite{mao98} based on the
nature of the host galaxies. 
Other recent analyses have also favored a
GRB redshift distribution which follows the strong redshift
evolution of the star formation rate \cite{sch99},\cite{frr00}.

One argument for strong redshift evolution for GRBs
which is independent of the SFR assumption has been put forward very 
recently. It is based on
correlating the time variability of GRB events with the absolute
luminosity of the events\cite{frr00}. The relationship is based on seven events
and has been extended to to a sample of 224 long, bright GRBs detected
by BATSE. The GRB rate was found to scale as $(1+z)^{3.6\pm0.3}$ which
is roughly consistent with the SFR for $z < 1.5$.  For higher $z$, the
SFR can not be accurately determined because it is strongly affected
by dust.  As we shall show in the next section, this uncertainty at such
high redshifts does not strongly
affect our result because contributions to the observed UHECR flux from such
redshifts are negligible.

\section{Calculations}
\label{calc.sec}

Energy losses of high energy protons from their creation to the
present epoch $(z = 0)$ are a result of cosmological redshifting and
by pair production ($p+\gamma\rightarrow p+e^+ + e^-$) and pion
production ($p+\gamma \rightarrow \pi +n$) through interactions with
cosmic background radiation (CBR) photons.  We shall assume for this
calculation a flat ($\Omega = 1$) Einstein-de Sitter universe with a
Hubble constant of H$_0$ = 70 km sec$^{-1}$ Mpc$^{-1}$ ($h=0.7$) and
taking $\Lambda = 0$.\footnote{Taking a $\Lambda$ = 0.7 model does not
significantly affect the results.} The
relationship between time and redshift is then given by
\begin{eqnarray}
  t= {2\over{3}} H_0^{-1}(1+z)^{-3/2},
  \label{eq1}
\end{eqnarray}
and the energy loss owing to redshifting is
\begin{eqnarray}
  -{1\over{E}}{dE\over{dt}} = H_{0}(1+z)^{3/2}.
  \label{eq2}
\end{eqnarray}

Photopion production is the dominant interaction between UHECRs and
the CMB for energies $>$ 10$^{20}$ eV.  Below energies of 3 $\times$
10$^{19}$ eV, protons will lose energy mainly through pair production.
The combined energy loss rate for both pion and pair production for
protons in collisions with photons of the 2.7K background at the
present epoch ($z=0$) is defined as
\begin{eqnarray}
  {1\over{E}}{dE\over{dt}} \equiv \tau (E)^{-1}.
  \label{eq3}
\end{eqnarray}

The photomeson loss time is given by Stecker\cite{stk68} and 
that for pair production is given by Blumenthal\cite{blu70}. 
In order to compute the energy losses due to
pion and pair production occurring at a redshift $z$, one must account for the
increased photon densities and energies at larger $z$. The energy of
photons increases as $(1+z)$ and the photon density increases as
$(1+z)^3$.  The energy loss for photons for any epoch $z$ can then be
expressed as
\begin{eqnarray}
  {1\over{E}}{dE\over{dt}} = (1+z)^3\tau [(1+z)E]^{-1}.
  \label{eq4}
\end{eqnarray}
We can then calculate the energy E$_0$(z) at which a proton is created at
a redshift $z$ whose observed energy today is E.

In order to calculate the (number) flux of protons observed on earth of
energy $E$, we follow the formalism derived in Berenzinsky and
Grigor'eva \cite{bg88}. Because of their high energies, we will assume
that these protons will propagate along roughly straight lines from their
source to the observer, unaffected by small intergalactic magnetic fields.
(Dropping this assumption increases their propagation time and energy losses
so that it would only strengthen our argument.) 

The observed flux from a volume element of the
sphere $dV = R^3(z)r^2drd\Omega$ is
\begin{eqnarray}
  {dj\over{dE}}dE = {{F(E_0,z)dE_0n(z)dV}\over{(1+z)4\pi R_0^2r^2}}.
  \label{eq5}
\end{eqnarray}
where F(E$_0$,z) is the flux of particles emitted at $z$ of energy
E$_0$ and $n(z)$ is the density of GRBs as a function of redshift. 
We will take this function to be of the form
\begin{eqnarray}
 n(z) = (1+z)^{(3+q)}n(0).
  \label{eq6}
\end{eqnarray}
Furthermore with $R_0 = R(z)(1+z)$ and $R(z)dr = cdt$, changing time
to redshift dependence utilizing (\ref{eq1}), and by integrating
(\ref{eq5}) to the edge of the sphere ({\it i.e.} out to a maximum
redshift $z_{max}$) one obtains
\begin{eqnarray}
 j(E) = {3\over{8\pi}}R_0n(0)F(E)\int_{0}^{z_{max}}(1+z)^{(q-5/2)}
\left({E_0\over{E}}\right)^{-\alpha}{{dE_0}\over{dE}}dz.
  \label{eq7}
\end{eqnarray}
assuming a power-law spectrum of index $\alpha$ for F(E)
to be generated by the GRBs,
where F(E) is the total number of particles per second with energy E
observed.

Above $\sim 10^{19}$ eV, the Fly's Eye\cite{fly94} and AGASA
data\cite{agasa95}\cite{agasa98} are well fit by a spectrum with a
power law $\propto$ E$^{-2.75}$. 
The ``GZK cutoff energy'' for UHECRs versus redshift of origin is plotted 
in Figure 1. We have defined this as the energy at which the spectrum drops to
$1/e$ of its original value owing to photomeson production. It can be seen 
from Figure 1 that the protons of energy above ($> 10^{20}$ eV) 
can have originated no further than a redshift of $z
= .034$ or a distance of 130 Mpc. 

For the redshift distribution of GRBs, we have have considered
three scenarios and corresponding values for the evolution index, $q$. 
In the first two cases, we assume a value for $\alpha$ of 2.75 which fits the
data above 10$^{19}$eV. For the first case, we take an isotropic comoving 
source density
distribution corresponding to $q = 0$ out to a maximum redshift of
2.5.  In a second case, we take a redshift 
dependence proportional to the
star formation rate corresponding to $q = 3$ out to a redshift of 2.5.

For our third case, we take the redshift distribution for GRBs derived 
from the time variability-redshift relationship for GRBs\cite{mao98}.
This corresponds to taking $q = 3.6$ out to a redshift of 3.6. (It should be
noted that there is no observable contribution to the cosmic ray flux
from redshifts beyond 2.5 in any case.) For this case, we assume a flatter 
source spectrum with a smaller value for $\alpha$ of 2.35. This gives a better
fit to the data between 10$^{18}$ and 10$^{19}$ eV, but requires a larger 
energy input. (If we take $\alpha$ = 2.75 for this case, we get a very similar
result to the second case, but slightly lower fluxes.)

We have normalized our resulting spectra to the Fly's Eye data at $E =
10^{18}$ eV and calculated the corresponding UHECR spectra observed at
Earth under the assumptions for the three scenarios described above. The 
results of these scenarios are
plotted against the Fly's Eye and AGASA data in figure 2.  The failure
of all three scenarios to fit the observed observed spectrum,
particularly for energies above 10$^{20}$ eV, is evident. In the
scenarios where the source rate follows the star formation rate as a
function of redshift, the spectrum is significantly steepened from the
source spectrum above 10$^{18}$ eV, a situation which does not occur in
the case of a uniform distribution independent of redshift.

For the $\alpha$ = 2.75 cases, the required
energy input in cosmic rays above 10$^{19}$ eV is $2 \times 10^{45}$ erg 
Mpc$^{-3}$ yr$^{-1}$. This is two orders of magnitude greater than the 
energy release rate from GRBs in the 0.01-1 MeV range \cite{stk00}.
For the $\alpha = 2.35$ case, the discrepency increases 
to three orders of magnitude, given the required normalization and the
flatter source spectrum.

One may speculate that there are numerous nearby low-flux GRBs below the
{\it BATSE} threshold. However, {\it BATSE} has only detected one burst
which may have had an energy release significantly lower than $10^{50}$ erg, 
even this event being in doubt \cite{bir93}-\cite{pia99}. Thus, there is no
reason to believe that such a scenario could overcome the energy discrepency
shown above unless much more GRB energy is released in UHECRs than in 
low-energy $\gamma$-rays.  

Our calculated spectrum above $\sim 10^{20}$ eV may be underestimated owing
to the fact that fluctuations in the interaction rate from very nearby sources
can allow more protons to reach the Earth than our continuous-energy-loss
approximation indicates \cite{aha94}. However, even with the inclusion of 
this effect, the predicted energy fluxes from GRBs would be much lower 
than the observations indicate\cite{bla00}. With regard to fluctuations, it 
has been argued that a few nearby sources could produce the UHECRs 
\cite{mir96} but statistical studies rule this out \cite{dub00}.

\section{Conclusions}

In this paper, we have presented the predicted spectra of
UHECRs assuming that the UHECR sources are GRBs and that
their source density distribution follows the star formation rate. 
Our motivation is the recent redshift
studies of GRBs which suggest a strong redshift evolution for
these objects.  The results indicate that GRBs can not be
responsible for the UHECRs observed at Earth
because their redshift evolution leads to a predicted
spectrum which is well below the data, particularly for energies above 
$10^{20}$ eV. In fact, the ``effective GZK cutoff'' for UHECRs of cosmological 
GRB origin is found to be $\sim 3 \times 10^{19}$eV (see Figure 2).

It should be noted that, independent of spectral considerations, if the
GRB redshift distribution follows the star formation rate, GRBs can be
ruled out as candidates for the UHECRs based on energetics considerations. The
present star formation rate coupled with the mean energy release per
burst leads to an energy release rate per volume which is more than an
order of magnitude below the rate need to account for the UHECRs with energies
above $10^{20}$ eV\cite{stk00},\cite{dar00} and, as shown in the previous 
section, two to three orders of magnitude below the rate needed to account for
UHECRs above $10^{19}$ eV, contradicting the hypothesis advocated in Ref.
\cite{wax95b} and stengthening the conclusion of Ref. \cite{stk00}.

The recently launched High Energy Transient Explorer (HETE-2)
and the Swift satellite mission, projected to be launched in
2003, should eventually provide redshift information for thousands of bursts,
including the short duration bursts.
These missions therefore have the potential to conclusively exclude
gamma ray bursts as the source of the UHECRs, should the density
distribution of the bursts conclusively indicate redshift evolution.

Future and planned UHECR detectors such as the Pierre Auger Observatory,
and the space-based orbiting wide-angle lens collector (OWL) satellite
system should
provide the coverage area necessary to determine the true particle
spectrum above 10$^{20}$eV. Confirmation of the lack of a pronounced GZK 
cutoff and an event rate similar to that indicated by present
data would also spell the demise for a GRB origin of the UHECRs.

\begin{figure}
  \centerline{\psfig{figure=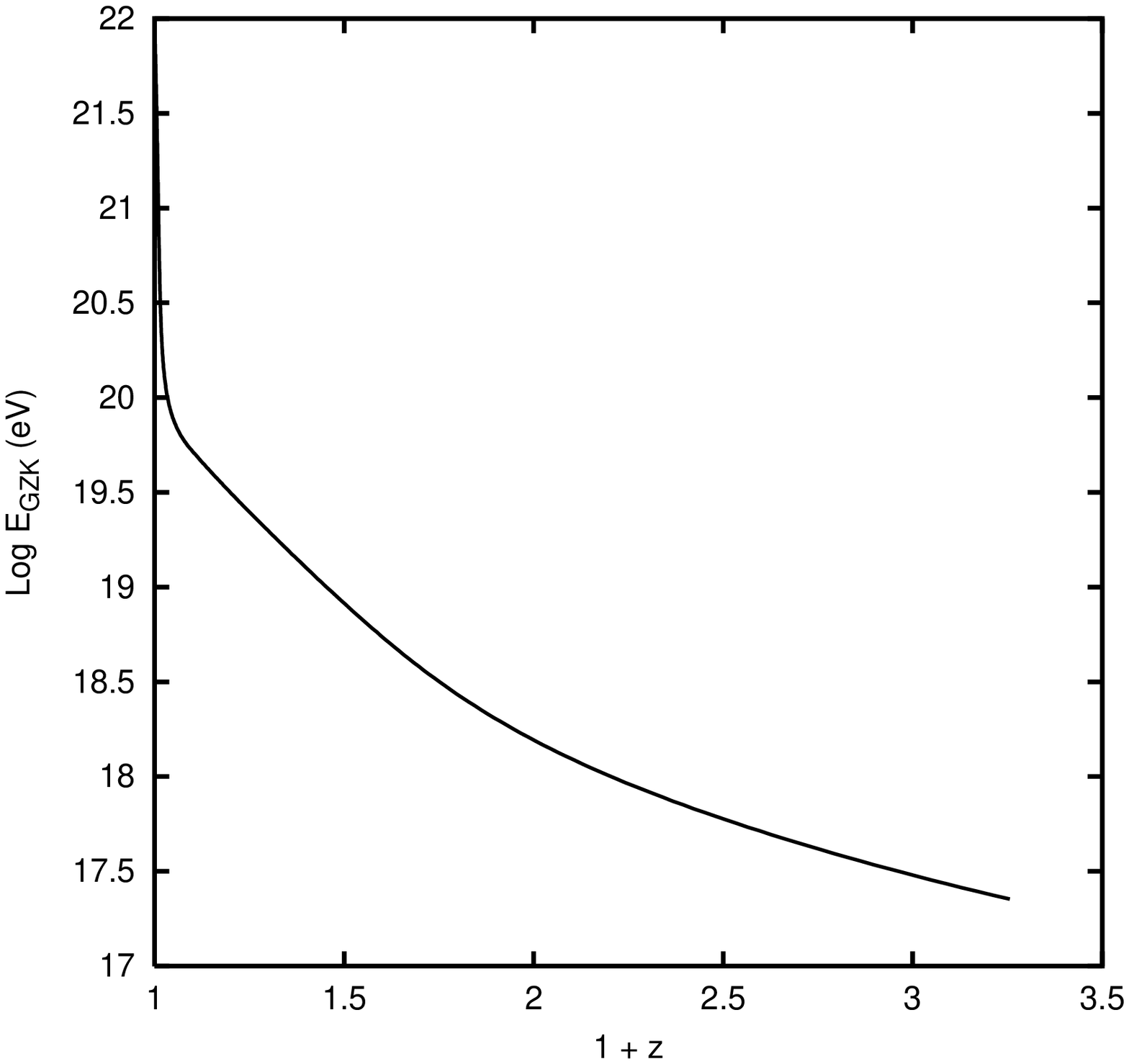,height=7.3in}}
  \caption{The maximum energy a proton should be observed with at Earth ({\it
    i.e.}, ``the GZK cutoff energy'') if it is produced at redshift $ z$.}
\end{figure}
\begin{figure}
  \centerline{\psfig{figure=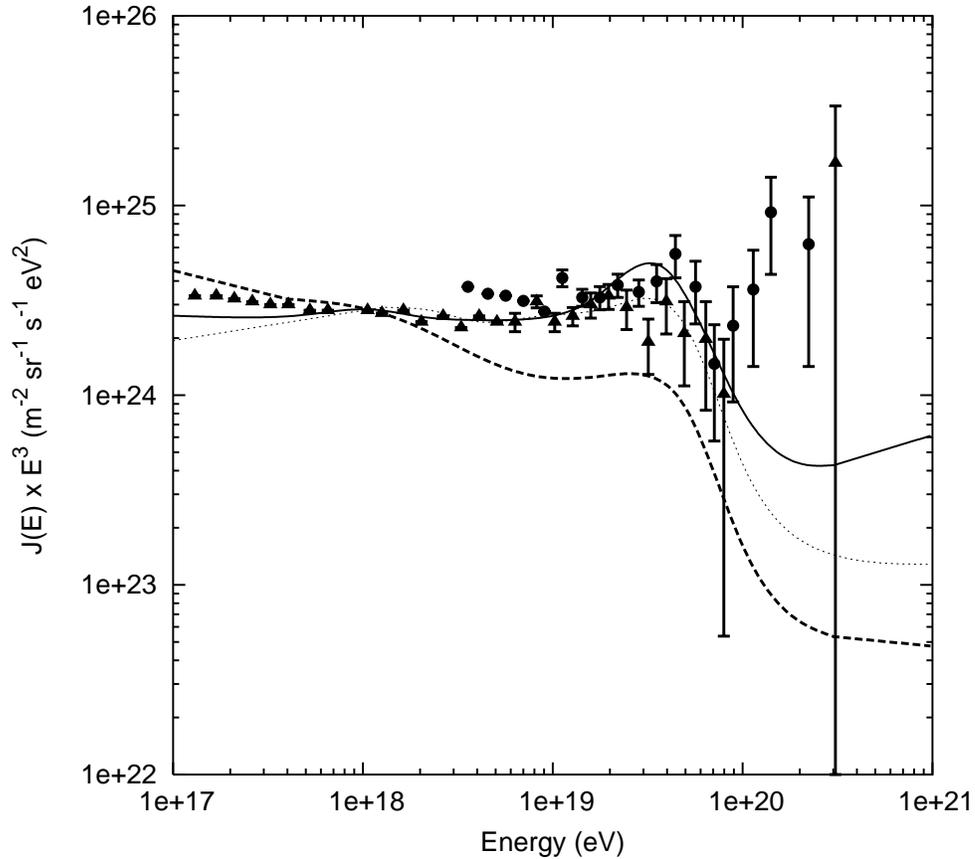,height=7.3in}}
  \caption{The highest energy region of the cosmic ray spectrum as
    observed by the Fly's Eye (triangles) and AGASA (filled dots)
    experiments. The solid line is the expected spectrum from sources
    whose density distibution follows the strong redshift
    dependence of gamma ray bursts derived in \protect\cite{frr00},
    {\it viz.}, ($q$ = 3.6, $z_{max}$=3.6) .  Also plotted are
    the expected spectra from sources whose density distribution
    evolves as the star formation rate \protect($q$ = 3.0, $z_{max}$=2.5) 
    (thick dashed line) and from
    sources whose distribution is independent of redshift \protect($q$ = 0.0)
    (dotted line). For the solid-line case, we have taken a source spectral
    index of 2.35; for the other two cases, an idex of 2.75 was chosen.}
\end{figure}

\end{document}